# Atmospheric Methane Removal as a Third Climate Intervention: Termination Risks and Air Pollutant Effects

Katsumasa Tanaka[1,2,✉], Weiwei Xiong[1], Didier A. Hauglustaine[1], Daniel J.A. Johansson[3], Nico Bauer[4], Philippe Bousquet[1], Philippe Ciais[1], Renaud de Richter[5], Marianne T. Lund[6], Ragnhild B. Skeie[6], Eric Zusman[7]

[1] Laboratoire des Sciences du Climat et de l'Environnement (LSCE), IPSL, CEA-CNRS-UVSQ, Université Paris-Saclay, Gif-sur-Yvette, France

[2] Earth System Division, National Institute for Environmental Studies (NIES), Tsukuba, Japan

[3] Division of Physical Resource Theory, Department of Space, Earth and Environment, Chalmers University of Technology, Gothenburg, Sweden

[4] Potsdam Institute for Climate Impact Research (PIK), Potsdam, Germany

[5] Tour-Solaire, Montpellier, France

[6] CICERO Center for International Climate Research, Oslo, Norway

[7] Institute for Global Environmental Strategies (IGES), Hayama, Japan

✉ Corresponding author: katsumasa.tanaka@lsce.ipsl.fr

**Abstract**

Atmospheric Methane Removal (AMR) is a third class of climate intervention, along with Carbon Dioxide Removal (CDR) and Solar Radiation Management (SRM). We show that, unlike CDR, the avoided warming by AMR is not durable due to methane's short atmospheric lifetime, although its temperature rebound upon termination is less abrupt than that of SRM. AMR's impact on tropospheric ozone can be further modulated by background pollutant levels.

**Main text**

Global long-term average surface temperatures are projected to exceed the 1.5 °C warming target set by the Paris Agreement within the coming years[1,2], making overshoot strategies increasingly important. There is a growing emission gap between current greenhouse gas (GHG) mitigation policies and the objectives of the Paris Agreement[3]. While deep emission reductions remain paramount, atmospheric removals become increasingly critical with time to complement emission reductions in achieving net-zero targets[4]. Carbon dioxide ($CO_2$) removals (CDRs) – technological and natural approaches that remove and durably store $CO_2$ from the atmosphere – are essential for returning below the 1.5 °C level in the long term; however, their actual implementation remains very limited[5]. The possibility of shaving the peak temperature with Solar Radiation Management (SRM) – artificial approaches that reduce the incoming surface solar radiation – is actively being discussed, but the climate and environmental uncertainties[6] associated with it, as well as legal and ethical issues, raise deep skepticism about its application[7,8].

This situation underscores the urgent need to explore a largely untapped opportunity: atmospheric methane ($CH_4$) removal (AMR). AMR, along with CDR and SRM, forms a *trio* of climate interventions – alternative mitigation approaches designed to supplement traditional emission reductions. AMR technologies, which oxidize or destroy atmospheric $CH_4$, are emerging as a potentially critical option, as recently highlighted by the US National Academies[9]. The concept of AMR was first proposed in 2010[10]. Since then, the research field has expanded, marked by a series of perspectives[11,12], modeling studies[13–15], and small-scale experimental and engineering projects[16,17].

$CH_4$ is a potent, short-lived GHG, currently the second largest radiative forcing agent after $CO_2$, with concentrations 2.6 times higher than pre-industrial levels[18]. While $CH_4$ has about 80 times higher radiative efficiency per unit mass than $CO_2$[19], $CH_4$ has a much shorter atmospheric perturbation time[20] of 11.8 ± 1.8 years[19] (versus $CO_2$ staying in the atmosphere over centuries to millennia). These differences in gas characteristics are a crucial factor to consider when developing mitigation strategies[21]. While $CO_2$ mitigation is the most important long-term solution to the climate problem, near-term $CH_4$ mitigation is effective for lowering the rate of warming and could limit an overshoot of the 1.5 °C warming target[22]. $CH_4$ mitigation can also lower tropospheric ozone levels and improve air quality, which would further reduce negative impacts on human health and ecosystems[23]. Currently, about 160 countries have signed up for the Global Methane Pledge (covering approximately 50% of anthropogenic $CH_4$ emissions). The signatories are committed to reducing global $CH_4$ emissions by 30% below 2020 levels by 2030, with a focus on low-cost mitigation opportunities in the oil and gas sector[24]. On the other hand, there are substantial $CH_4$ emissions, particularly from agriculture[25], that remain challenging to abate due to deeply rooted socio-economic factors, including dietary choices[26]. Such $CH_4$ emissions will not cause further warming unless they increase further (i.e., rapid equilibration due to methane's short lifetime[21]), but AMR could be useful for offsetting emissions from such hard-to-abate sectors, and would reduce the global warming contribution from $CH_4$. Finally, AMR could be useful for balancing the potential increase in natural $CH_4$ emissions from climate-sensitive sources, such as wetlands (including peatlands), permafrost, and methane hydrates.

A variety of AMR technologies (Box 1) are under consideration, each presenting its own advantages and disadvantages[9,27,28]. Associated costs are currently highly uncertain, as these technologies are in the early stages of development[29]. AMR can indirectly exert positive or negative impacts on human health and ecosystems by altering surface ozone, hydroxyl radical (OH), and PM2.5 concentrations[14,30], while certain CDR and SRM methods also affect air quality by releasing biogenic volatile organic compounds (VOC) and aerosol precursors[31–33]. A critical consideration that has received insufficient attention, however, is the *potential risks of terminating AMR*, for example, due to unforeseen side effects or unexpected political and economic crises. Drawing a parallel with the termination effect of SRM[34], an abrupt cessation of AMR could rapidly unmask warming, with unintended impacts on air quality. Arguably, climate interventions that require continuous active deployment to maintain their climate benefits (e.g., atmospheric oxidation enhancement and stratospheric aerosol injection) may be more vulnerable to abrupt termination than approaches based on biological carbon storage (e.g., afforestation). For the latter, however, the risk lies more in non-permanence[35] than in termination.

**Box 1. A summary of AMR technologies under consideration**

- *Methane reactors*: This method uses a photocatalytic or thermocatalytic reactor to break down the $CH_4$ molecule. Given the low ambient $CH_4$ concentration (approximately 2ppm for $CH_4$, in comparison to 420ppm for $CO_2$), the use of this technology would be best suited to closed systems with high $CH_4$ concentrations rather than processing a vast amount of air mass.
- *Atmospheric oxidation enhancement*: This method aims to accelerate the breakdown of $CH_4$ in open systems by increasing the concentration of atmospheric oxidants, such as hydroxyl radical (OH) and chlorine (Cl) through releasing hydrogen peroxide or iron salt aerosols. Because this approach directly alters atmospheric chemistry, a careful assessment of its side effects on human health and ecosystems is crucial.
- *Ecosystem uptake enhancement*: This biological approach seeks to increase the $CH_4$ consumption of microbes in the soils or by aboveground vegetation. The potential to engineer methanotrophs to enhance their ability to oxidize methane with higher growth rates requires research.
- *Surface treatment*: This technology can break down $CH_4$ using coating on panels, rooftops, or other surfaces exposed to sunlight using photocatalysts. This approach has been proposed to improve air quality in polluted areas, but its potential as a $CH_4$ removal requires further research.

Using a simplified global scenario-based approach, we illustrate the potential magnitude and duration of the effects of terminating AMR relative to CDR and SRM. We first define a climate intervention scenario (reference scenario), in which AMR, CDR, and SRM are deployed concurrently from 2040 onwards (Fig. 1, Methods). To ensure comparability, each intervention is introduced so as to have an identical forcing profile (Fig. 1i). To simulate termination risks, each intervention is individually phased out over five years, beginning in 2050, 2060, or 2070. Our model represents deployments and terminations generically by emissions removed (for AMR and CDR) or negative radiative forcing (for SRM) without accounting for further processes related to specific technologies.

Using the reduced-complexity climate model ACC2[4,36] (Methods), we calculated the global temperature effects of deploying and terminating each intervention (Fig. 1a). AMR, CDR, and SRM deployments provide an almost identical cooling effect, confirming the relatively weak state dependence and moderate feedbacks within this scenario framework. The terminations, however, reveal distinct temperature responses. While all terminations lead to warming relative to the levels without terminations (i.e., temperature rebound), SRM termination causes the most rapid rebound. The CDR termination results in a moderate rebound and, unlike the other two, does not result in a full return due to the assumed permanence of $CO_2$ removal. On the other hand, the AMR termination results in a full return like the SRM termination, indicating that the cooling effect of AMR is not permanent, although its termination rebound is less rapid than that of SRM. Importantly, AMR termination has impacts on atmospheric chemistry, decreasing OH concentrations over about a decade after termination (Fig. 1e), which increases the $CH_4$ lifetime (Fig. 1d), tropospheric ozone forcing (Fig. 1h), and stratospheric water vapor forcing (not shown; an increase $< 0.01$ W/m$^2$).

The temperature and air quality impacts of AMR deployment and termination are dependent on background pollutants[37,38] (Fig. 2). Air pollutants considered here ($NO_x$, CO, and VOC) are assumed to follow SSP5-3.4-OS by default (Fig. S1). They do not directly affect the temperature, but they indirectly affect it through changes in $CH_4$ and tropospheric ozone concentrations. With $NO_x$ following a lower emission scenario (SSP1-2.6), lower $NO_x$ emissions enhance $CH_4$ concentrations (Fig. 2b) by limiting the OH recycling mechanisms and reducing the tropospheric ozone production (Fig. 2c); the latter further reduces OH production through limiting ozone photolysis. The two opposing effects – $CH_4$ enhancement and ozone suppression – result in a small net warming (Fig. 2a). These chemical impacts affect the scenarios with and without AMR termination almost equally (Figs. S2 and S3), without causing visible changes in temperature rebounds (the lines overlap in the Fig. 2a inset). Conversely, with $NO_x$ following a higher scenario (SSP5-8.5), the chemical impacts are opposite; however, the net result is a larger warming driven by strong ozone enhancement.

Lower CO and VOC emissions (treated together due to their similar chemical impacts) lead to a small decrease in $CH_4$ concentrations because they compete less with $CH_4$ for OH oxidation, making OH more available for $CH_4$ (sink effect). They also slightly reduce tropospheric ozone (by reducing the conversion of NO to $NO_2$) and dampen OH production (source effect). The source effect is outweighed by the sink effect, resulting in a net decrease in $CH_4$. The combined decrease in $CH_4$ and tropospheric ozone lowers the warming.

With all three pollutants following the low scenario, the net result is a slight cooling, accompanied by lower ozone forcing early this century and nearly unchanged $CH_4$ levels. In contrast, with all pollutants following the high scenario, the net result is a slight warming, accompanied by increased ozone and decreased $CH_4$ later this century. In both scenarios, however, the temperature rebound remains largely unaffected.

In our reduced-complexity modeling approach, the temperature and chemical effects caused by pollutants are largely linear and independent of $CH_4$ concentrations (Figs. S2-S5). It is, however, important to emphasize that AMR deployment and termination can have strong nonlinear impacts on the full chemistry system including sulfate and nitrate aerosols[14,30], implying consequences for surface air quality and potential impacts on human health and ecosystems. These impacts also depend on regional and seasonal factors, requiring investigation using multiple spatially-explicit Chemistry Transport Models resolving detailed process-level nonlinear interactions[37,39,40]. Such models are also better equipped to resolve surface ozone than reduced-complexity models (Methods). The chemical impacts of AMR remain largely unexplored in comparison to those of CDR and SRM[31–33]. Furthermore, as Earth system models begin to include an interactive methane cycle, these models could be useful for understanding various impacts of AMR[13].

Overall, with its distinct climate and air quality effects, AMR warrants further investigation for its potentially unique and important role in the climate intervention trio. It could conceivably serve as *an alternative to the riskier SRM option*, which carries various side effects, including impacts on the hydrological cycle[8]. This raises a question for future research:

to what extent can AMR reduce the potential reliance on solar geoengineering? Given its rapid response to temperature overshoot, AMR might avoid the termination risks and side effects associated with SRM, albeit with currently uncertain implications for air quality. Addressing this inquiry requires detailed evaluations of individual AMR and SRM methods – particularly their environmental side effects – supported by spatially explicit and process-based modeling approaches.

While better understanding of these physical and chemical uncertainties is crucial, realizing AMR's full potential requires more than scientific understanding and technological breakthroughs. A range of economic, legal, social, and institutional barriers – as is also the case with CDR and SRM – could significantly impede its adoption. Because AMR targets $CH_4$, which behaves fundamentally differently from $CO_2$, it raises unique challenges regarding how to integrate AMR with existing mitigation frameworks[41], such as emission offset markets[42]. It also requires societal engagement to communicate the distinct nature of $CO_2$ versus $CH_4$ removals[43]. Hence, future research needs to extend beyond technological feasibility and integrate Earth system modeling with economics and social sciences to create an enabling environment for potential implementation.

## Methods

SSP5-3.4-OS scenario and AMR, CDR, and SRM profile

To maintain consistency with literature, we developed our climate intervention scenarios based on an interpretation of the SSP5-3.4-OS scenario, which was derived from the REMIND-MAgPIE[44] Integrated Assessment Model as part of CMIP6[45] (Fig. 1). This overshoot scenario follows the very high SSP5-8.5 until 2040 and then branches off in 2040, after which drastic mitigation is assumed. SSP5-3.4-OS achieves net-zero $CO_2$ emissions by late 2060s (and approximately −20 $GtCO_2$/year emissions by 2100) and a 77% $CH_4$ emission reduction by 2070 relative to SSP5-8.5 (Fig. S1).

We determined the deployment profiles of AMR, CDR, and SRM as explained below. First, it is important to note that the CDR and SRM deployment profiles discussed here are, by design, equivalent to the AMR deployment profile; this discussion does not refer to the underlying CDR profile that shapes the background temperature overshoot.

AMR deployment profile

Of the three major anthropogenic $CH_4$ sources (energy, agriculture, and waste sectors) (Fig. S6), we assume that $CH_4$ emission reductions from the "hard-to-abate" agriculture and waste sectors in SSP5-3.4-OS (relative to SSP5-8.5) will be achieved through AMR starting in 2040 (red shaded area, Fig. 1b). Reductions from the fossil fuel sector are assumed to occur through traditional mitigation (not shown in Fig. 1). For simplicity, we do not consider applications of AMR to natural $CH_4$ sources in our scenarios, although we acknowledge its potential utility in managing climate-driven emissions from natural sources[18].

CDR deployment profile

We employed two different methods to determine a CDR deployment profile *equivalent* to the AMR profile.

First, our primary method, termed *forcing equivalence method (default, model-based)*, equates CDR and AMR deployments by radiative forcing. We numerically optimized annual $CO_2$ removals starting in 2040 (green shaded area, Fig. 1f) such that the resulting radiative forcing of CDR matches that of AMR (green and red shaded areas, respectively, Fig. 1i). Through this method, the resulting CDR profile reflects not only AMR's direct effect on methane but also its indirect effects on tropospheric ozone and stratospheric water vapor. The resulting $CO_2$ removal decreases later in this century because the $CH_4$ removal becomes nearly stable (red shaded area, Fig. 1b); once the atmospheric $CH_4$ response equilibrates (due to methane's short lifetime), it generates little additional negative forcing[21], reducing the required rate of $CO_2$ removal. Furthermore, the ratio of $CH_4$ removal to $CO_2$ removal (Fig. S7) – referred to as the Forcing Equivalent Index[46–48] – is strongly time-dependent. It is proximate to the 20-year Global Warming Potential (GWP20) for the first few decades but declines towards GWP100 and the 100-year Global Temperature change Potential (GTP100) later in the century. The small negative value at the end of the simulation (2120) reflects the decline in AMR deployment; the resulting positive forcing requires negative CDR (or positive emissions) to maintain forcing equivalence.

Second, our alternative method, termed *emission equivalence method (policy-oriented, metric-based)*, equates CDR and AMR deployments by mass using three GHG emission metrics[49]: GWP100, GWP20, and GTP100. Metric values for $CH_4$ (28, 84, and 4, respectively) were taken from the Intergovernmental Panel on Climate Change (IPCC) Fifth Assessment Report (AR5), as adopted for use under the Paris Agreement[36]. In the GWP100 case – the default metric of the Paris Agreement – the annual $CO_2$ removal mass equals the $CH_4$ removal mass multiplied by 28. GWP20 and GTP100 serve as high- and low-sensitivity cases. The emission equivalence method yields $CO_2$ removal profiles that differ significantly from the profile derived using the forcing equivalence method (Fig. S8). The CDR profile based on GWP20 closely follows the default forcing-equivalence profile for the first 30 to 40 years but subsequently overestimates it. In contrast, the profiles based on GWP100 and GTP100 consistently underestimate the default CDR profile.

SRM deployment profile

SRM deployment (represented by radiative forcing) was assumed to be equivalent to the AMR forcing profile (yellow shaded area, Fig. 1i). This SRM component was added to our interpreted SSP5-3.4-OS, which already incorporates AMR and CDR, to create a reference scenario including all three interventions (black lines). Consequently, while AMR and CDR deployments are depicted above the SSP5-3.4-OS level in Fig. 1, SRM deployment is indicated below that level.

### AMR, CDR, and SRM termination profile

The termination scenarios assume a linear phase-out of the respective interventions over five years (red, green, and yellow lines, respectively, Fig. 1b), beginning in 2050, 2060, or 2070. Specifically, the CDR termination scenario involves deploying and terminating CDR; the AMR termination scenario involves deploying both CDR and AMR, but terminating only AMR; and the SRM termination scenario involves deploying CDR, AMR, and SRM, but terminating only SRM. The termination start years are illustrative and chosen to explore climate and chemistry implications over subsequent decades. The five-year phase-out is similarly illustrative and represents a rapid cessation of interventions. We do not model specific intervention technologies and do not consider energy and material requirements. Among other factors, our scenarios do not account for i) the non-permanence of CDR technologies (e.g., afforestation and reforestation), ii) the chemical implications of AMR technologies that directly alter OH or Cl concentrations via hydrogen peroxide dispersal and iron salt aerosols, respectively, and iii) the impact of SRM on the carbon cycle through changes in direct and diffuse radiation.

### ACC2 model description

ACC2 is a reduced-complexity global climate model comprising (i) carbon cycle, (ii) atmospheric chemistry, and (iii) physical climate modules; a full model description is provided in ref[50]. The mitigation module – a component in ACC2 used to calculate mitigation costs in cost-effective pathways[51] – was not used in this study. ACC2 belongs to the class of reduced-complexity models also known as climate emulators[52,53], which are calibrated to historical observations and/or the output of more complex models, primarily aiming to represent Earth system dynamics at the global-annual-mean level without internal variability. Climate emulators are often used for policy applications and assessments.

The global carbon cycle is represented by a box model: four boxes representing the coupled atmosphere-ocean and another four for the land. Saturation of ocean $CO_2$ uptake with rising atmospheric $CO_2$ concentrations is modeled through the thermodynamic equilibrium of ocean carbonate species. $CO_2$ fertilization of the land biosphere is parameterized by a commonly used beta factor. Climate-carbon feedbacks are parameterized by a Q10 factor. ACC2 covers a comprehensive set of direct and indirect climate forcers: $CO_2$, $CH_4$, $N_2O$, ozone, $SF_6$, 29 species of halocarbons, OH, $NO_x$, CO, VOC, aerosols (both radiative and cloud interactions), and stratospheric $H_2O$. Importantly, each forcing term is calculated separately without any gas aggregation using metrics such as GWP100. The equilibrium climate sensitivity is assumed at 3 °C. Other uncertain parameters are optimized by combining an inverse estimation approach with historical data[57]. In this paper, we present projections based on the best parameter estimates for illustrative purposes. The model is written in GAMS and numerically solved using CONOPT4.

The atmospheric chemistry is highly parameterized and based on sensitivity analyses using several Chemistry Transport Models (Table 4.11 of ref[54]), as described in Section 2.2.2 of ref[50] and also adopted by other reduced-complexity climate models[55,56]. Our parameterization represents interactions between $CH_4$, OH, ozone, and pollutants without dependence on temperature, humidity, or other factors[57]. ACC2 considers three sinks for $CH_4$: depletion by OH oxidation, stratospheric loss, and soil sink, with the lifetime of 8.5 years (2000 value based on inversion[58]), 120 years, and 160 years, respectively. The $CH_4$ lifetime with respect to OH oxidation is inversely related to relative OH concentrations (2000 basis). The total atmospheric $CH_4$ lifetime (Fig. 1d) is 7.6 years (2008-2017 mean), which is shorter than the 1σ range of 9.1±0.9 years of IPCC AR6[59] (Table 6.2). The tropospheric ozone forcing scales linearly with the tropospheric ozone column burden, given in Dobson Units (DU), with a radiative efficiency of 0.042 $W/m^2/DU$[50]. Unlike well-mixed gases, such as $CO_2$ and $CH_4$, ozone exhibits a highly variable vertical distribution with a short atmospheric lifetime (from a few hours up to a few months depending on the location)[59]. Because our modeling approach is not designed to simulate surface ozone directly, tropospheric ozone forcing is interpreted as a proxy for surface ozone and air quality. This approximation is supported by established literature demonstrating that methane-driven changes in the tropospheric ozone column are indicative of changes in surface ozone concentrations[23,37,59,60]. Furthermore, it is considered that the change in methane concentrations has a positive relationship with the change in surface ozone concentrations. Our approach does not consider the temperature feedback on water vapor concentrations, which would also affect OH production[61] and thus $CH_4$ lifetime.

**Data Availability:** Data used to generate the figures in the paper are available on Zenodo with https://doi.org/10.5281/zenodo.18488977

**Code Availability:** Code used to generate the figures in the paper are available on Zenodo with https://doi.org/10.5281/zenodo.18488977

**Acknowledgments:** This work was supported by the European Union's Horizon Europe research and innovation program under Grant Agreements N° 101056939 (RESCUE – Response of the Earth System to overshoot, Climate neUtrality and negative Emissions) and N° 101081193 (OptimESM – Optimal High Resolution Earth System Models for Exploring Future Climate Changes). We thank Olivier Boucher and Yann Gaucher for their perspectives and insights, which helped shape this work.

**Author contributions:** K.T. conceived and led this study. K.T., D.H., and D.J. developed the scenarios and analytical framework, with subsequent contributions from all other co-authors (W.X., N.B., P.B., P.C., R.R., M.L., R.S., and E.Z.). K.T. and W.X. performed the ACC2 simulations. W.X. prepared the figures with input from K.T. K.T. wrote the original draft, and all authors contributed to the review and editing of the manuscript.

**Competing interests:** The authors declare no competing financial or non-financial interests.

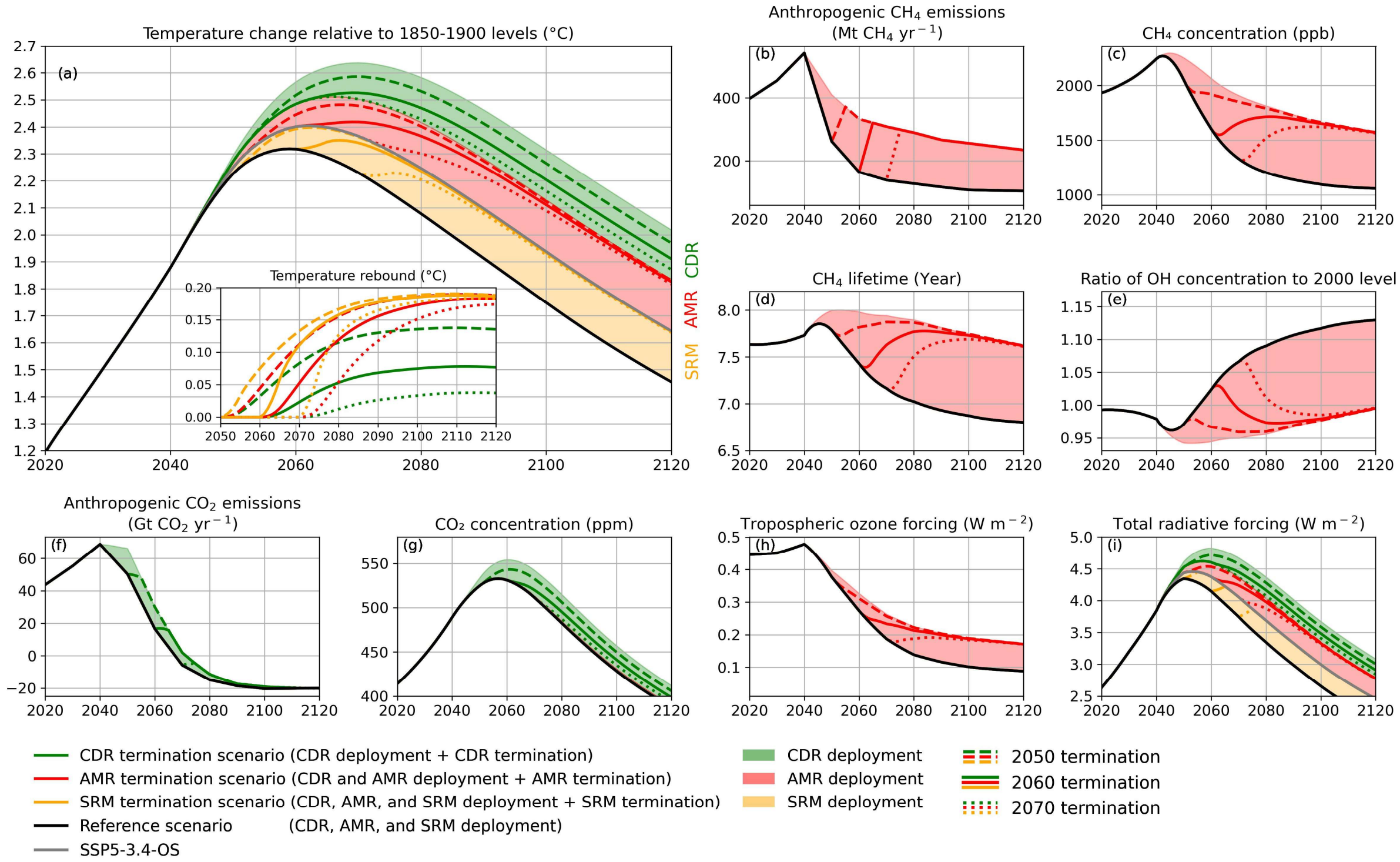

**Fig. 1 | Deployments and terminations of the trio of climate interventions: AMR, CDR, and SRM.** (a) Global-mean temperature change relative to 1850–1900 levels (°C); (b) Anthropogenic $CH_4$ emissions ($MtCH_4$/yr); (c) Atmospheric $CH_4$ concentration (ppb); (d) Atmospheric $CH_4$ lifetime (yr); (e) Ratio of OH concentration to 2000 levels; (f) Anthropogenic $CO_2$ emissions ($GtCO_2$/yr); (g) Atmospheric $CO_2$ concentration (ppm); (h) Tropospheric ozone forcing ($W/m^2$); (i) Total radiative forcing ($W/m^2$). The black line represents the reference scenario, in which AMR, CDR, and SRM are combined and deployed from 2040 onwards. The red, green, and yellow shaded areas are the effects of AMR, CDR, and SRM deployments, respectively. CDR and SRM deployments shown here are, by design, equivalent to the AMR deployment in terms of forcing; the larger underlying CDR driving the background temperature overshoot pathway is not explicitly indicated. The red, green, and yellow lines are the projections of AMR, CDR, and SRM terminations, respectively, beginning in 2050, 2060, or 2070 (dashed, solid, and dotted lines, respectively). The inset within panel (a) indicates the *temperature rebound*, that is, the temperature change after the termination of each intervention relative to the warming level without the termination (e.g., for AMR termination, the temperature change relative to the gray line). The atmospheric $CH_4$ lifetime (panel (d)) is for all sinks in our model (OH depletion, soil uptake, and stratospheric loss). In panels (b) to (h), the black line overlaps exactly with the gray line, except for panel (g) with a very small difference due to carbon cycle feedback. The gray line (SSP5-3.4-OS) does not exactly reach 3.4 $W/m^2$ in 2100, despite what its name suggests, because of the difference between ACC2 and the reduced-complexity climate model used to develop SSP5-3.4-OS. Projections are based on best parameter estimates for illustrative purposes.

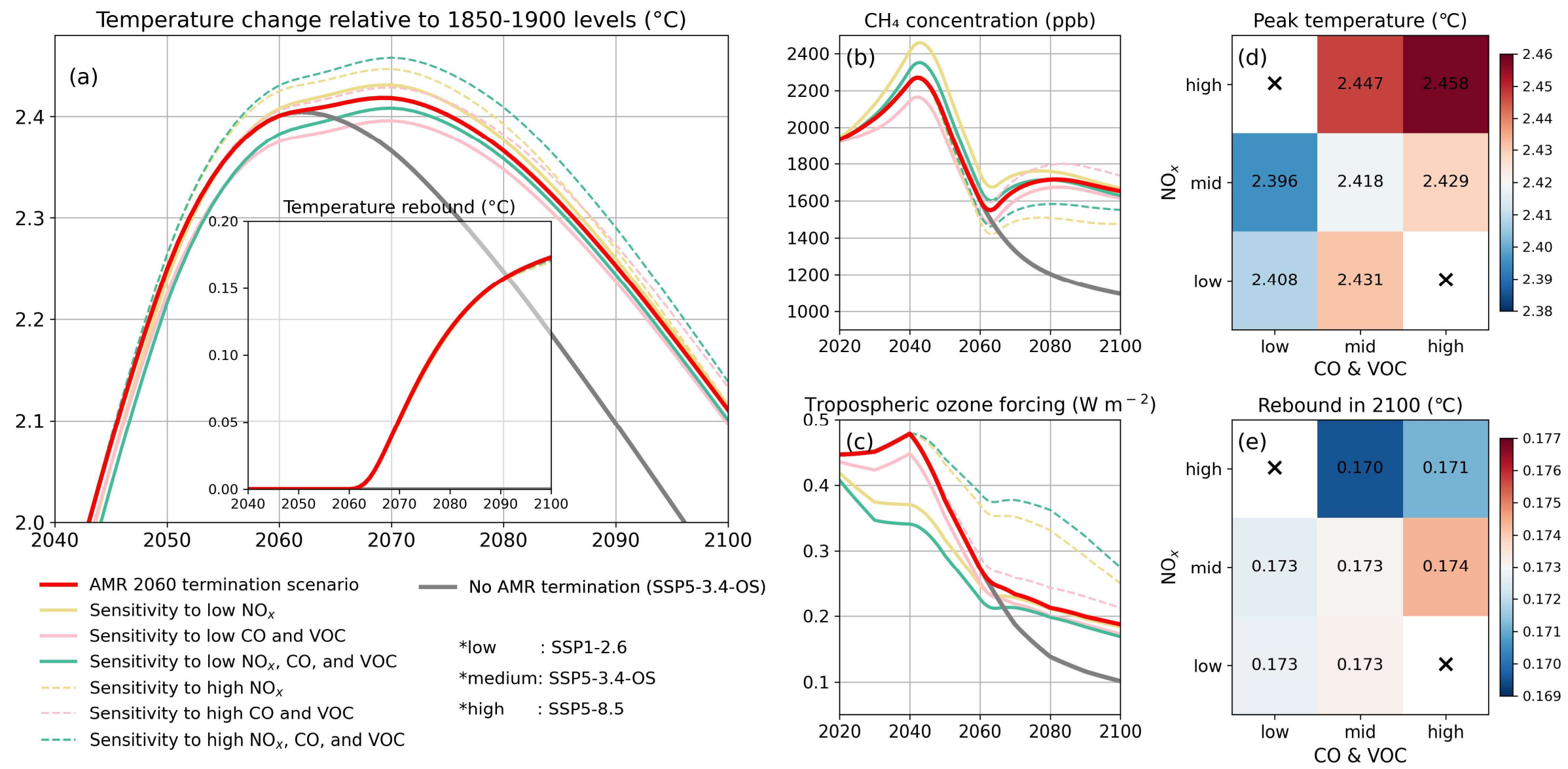

**Fig. 2 | AMR deployment and 2060 termination effects under different background pollutant emissions.** (a) Global-mean temperature change relative to 1850–1900 levels (°C) with inset showing temperature rebound (°C); (b) Atmospheric $CH_4$ concentration (ppb); (c) Tropospheric ozone forcing (W/m$^2$); (d) Peak temperature (°C) under different background pollutant emissions; (e) Temperature rebound in 2100 (°C) under different background pollutant emissions. This figure shows the sensitivity of temperature and air quality (tropospheric ozone) impacts of the AMR 2060 termination scenario (red line) to variations in pollutant emission pathways ($NO_x$, CO, and VOC). Tropospheric ozone forcing is shown as a proxy for surface ozone and air quality (Methods). Default pollutant emissions follow SSP5-3.4-OS (Fig. S1). In the sensitivity cases (legend), the indicated pollutant follows either the low-emission SSP1-2.6 pathway (thick lines in panels (a), (b), and (c)) or the high-emission SSP5-8.5 pathway (thin lines in panels (a), (b), and (c)), while all other pollutants remain at SSP5-3.4-OS. The temperature rebound in the inset of panel (a) indicates the temperature difference due to the AMR termination *under consistent pollutant assumptions*; for example, the temperature rebound of the "Sensitivity to low $NO_x$" case represents the temperature difference between the AMR 2060 termination scenario and the no AMR termination scenario, both with $NO_x$ following SSP1-2.6 and CO and VOC following SSP5-3.4-OS (Figs. S2 and S3). The temperature rebounds of all sensitivity cases largely overlap with that of the default case due to the limited sensitivity to these variations. Panels (d) and (e) summarize the main features of the sensitivity results. We excluded mixed cases (e.g., high $NO_x$ combined with low CO and VOC) to avoid strong internal inconsistency within the underlying pollutant scenarios.

# Atmospheric Methane Removal as a Third Climate Intervention: Termination Risks and Air Pollutant Effects

## – Supplementary Information –

Katsumasa Tanaka[1,2,✉], Weiwei Xiong[1], Didier A. Hauglustaine[1], Daniel J.A. Johansson[3], Nico Bauer[4], Philippe Bousquet[1], Philippe Ciais[1], Renaud de Richter[5], Marianne T. Lund[6], Ragnhild B. Skeie[6], Eric Zusman[7]

[1] Laboratoire des Sciences du Climat et de l'Environnement (LSCE), IPSL, CEA-CNRS-UVSQ, Université Paris-Saclay, Gif-sur-Yvette, France

[2] Earth System Division, National Institute for Environmental Studies (NIES), Tsukuba, Japan

[3] Division of Physical Resource Theory, Department of Space, Earth and Environment, Chalmers University of Technology, Gothenburg, Sweden

[4] Potsdam Institute for Climate Impact Research (PIK), Potsdam, Germany

[5] Tour-Solaire, Montpellier, France

[6] CICERO Center for International Climate Research, Oslo, Norway

[7] Institute for Global Environmental Strategies (IGES), Hayama, Japan

✉ Corresponding author: katsumasa.tanaka@lsce.ipsl.fr

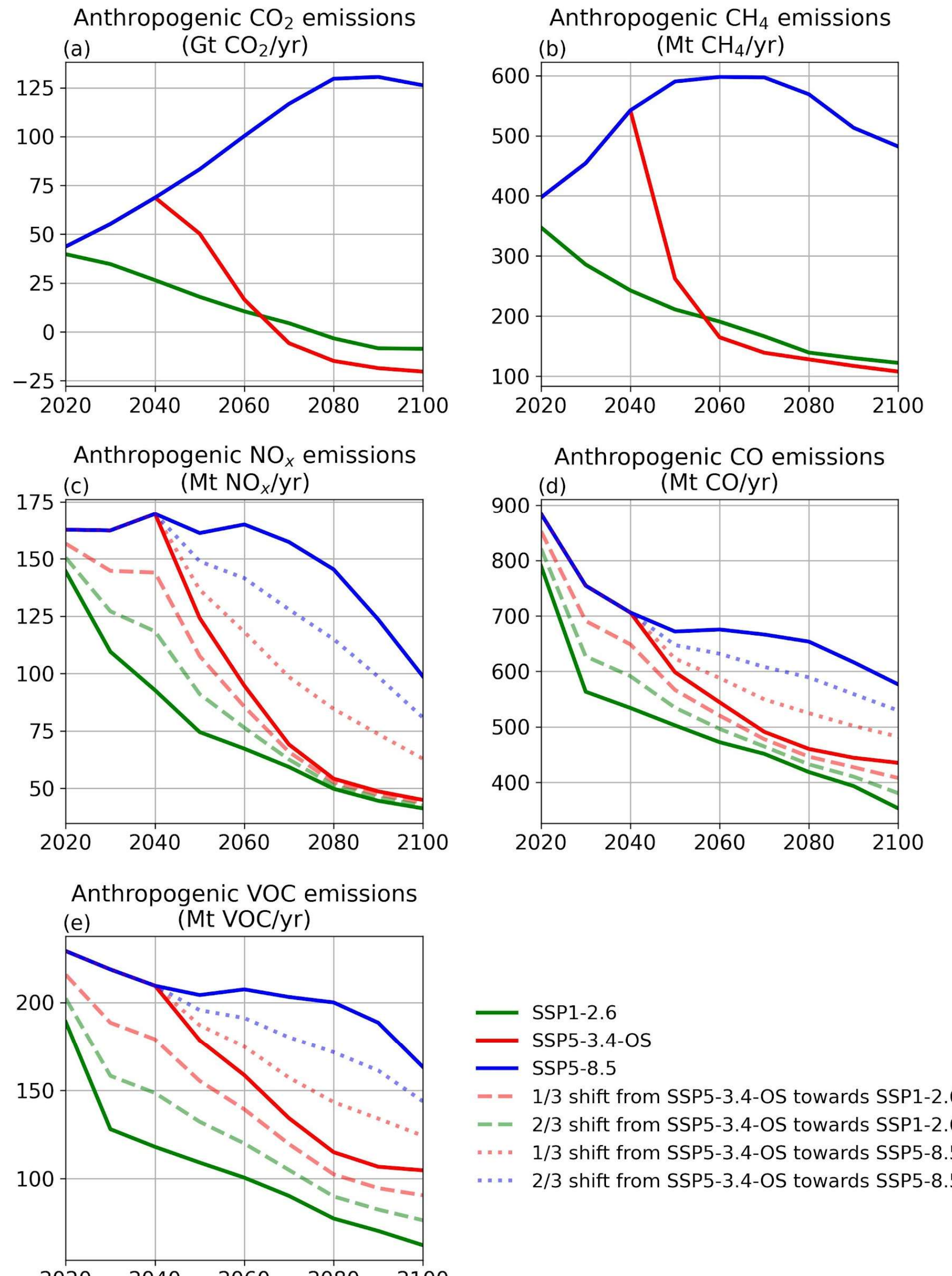

**Fig. S1 | $CO_2$, $CH_4$, and pollutant pathways of SSP1-2.6, SSP5-3.4-OS, and SSP5-8.5.** (a) Anthropogenic $CO_2$ emissions ($GtCO_2$/yr); (b) Anthropogenic $CH_4$ emissions ($MtCH_4$/yr); (c) Anthropogenic $NO_x$ emissions ($MtNO_x$/yr); (d) Anthropogenic CO emissions (MtCO/yr); (e) Anthropogenic VOC emissions (MtVOC/yr). Data are derived from the IIASA SSP Database v2.0 (https://tntcat.iiasa.ac.at/SspDb/dsd). The four interpolated scenarios are used in the sensitivity analysis presented in Figs. S4 and S5.

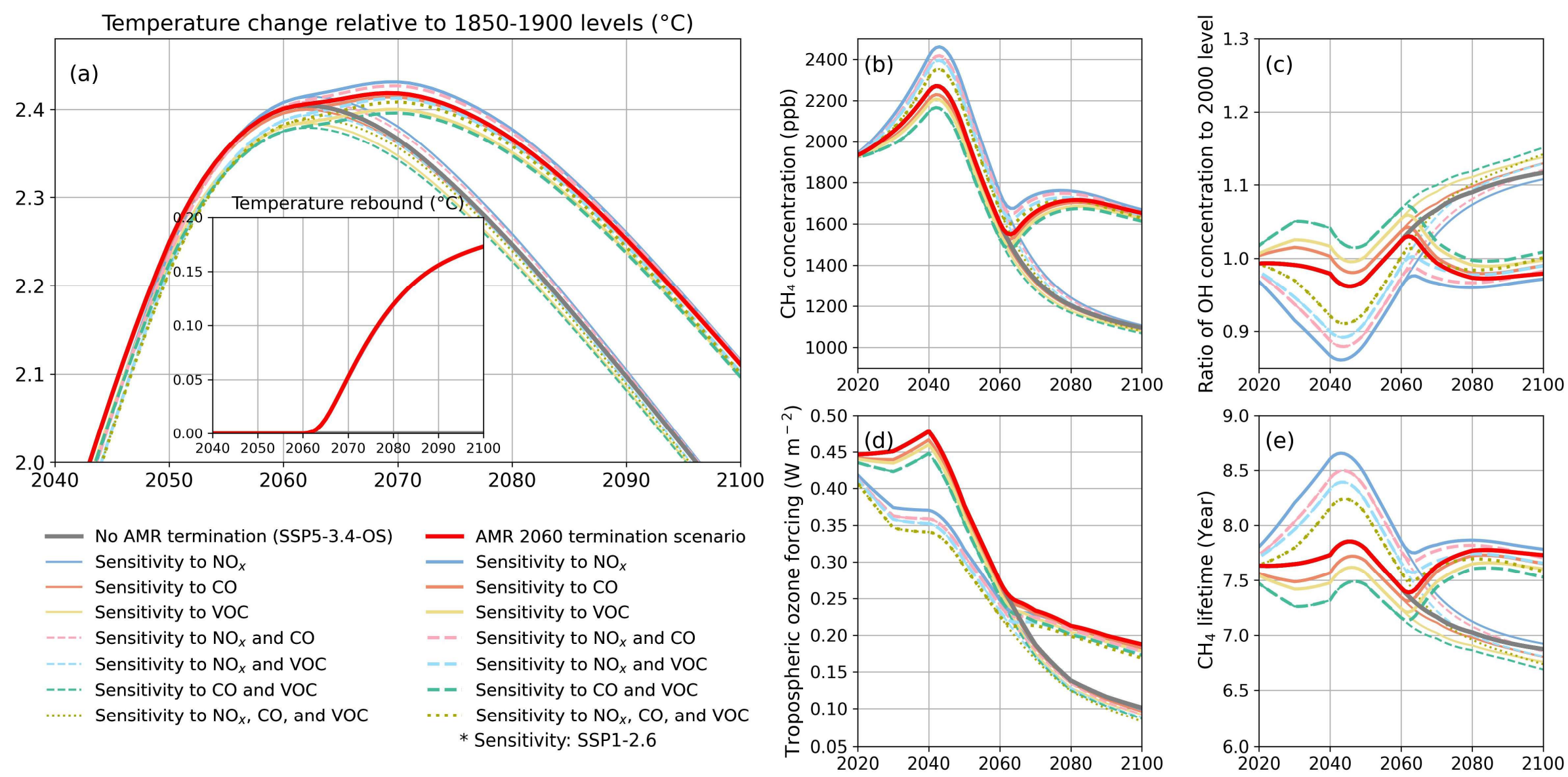

**Fig. S2 | Sensitivity of scenarios with and without AMR termination to background pollutant emissions at SSP1-2.6.** (a) Global-mean temperature change relative to 1850–1900 levels (°C) with inset showing temperature rebound (°C); (b) Atmospheric $CH_4$ concentration (ppb); (c) Ratio of OH concentration to 2000 levels; (d) Tropospheric ozone forcing ($W/m^2$); (e) Atmospheric $CH_4$ lifetime (yr). This figure complements Fig. 2 by extending the sensitivity analysis to the scenario without AMR termination and to all possible combinations of pollutants. It displays the response of the scenarios with AMR termination (red line) and without (gray line) to variations in pollutant emissions ($NO_x$, CO, and VOC). In these sensitivity cases, the indicated pollutant is assumed to follow SSP1-2.6. Line thickness distinguishes the base scenario: thick lines represent sensitivities applied to the scenario with AMR termination, while thin lines represent those applied to the scenario without AMR termination. For example, the thick line for "Sensitivity to $NO_x$" indicates an AMR termination scenario where $NO_x$ emissions follow SSP1-2.6, with CO and VOC following SSP5-3.4-OS.

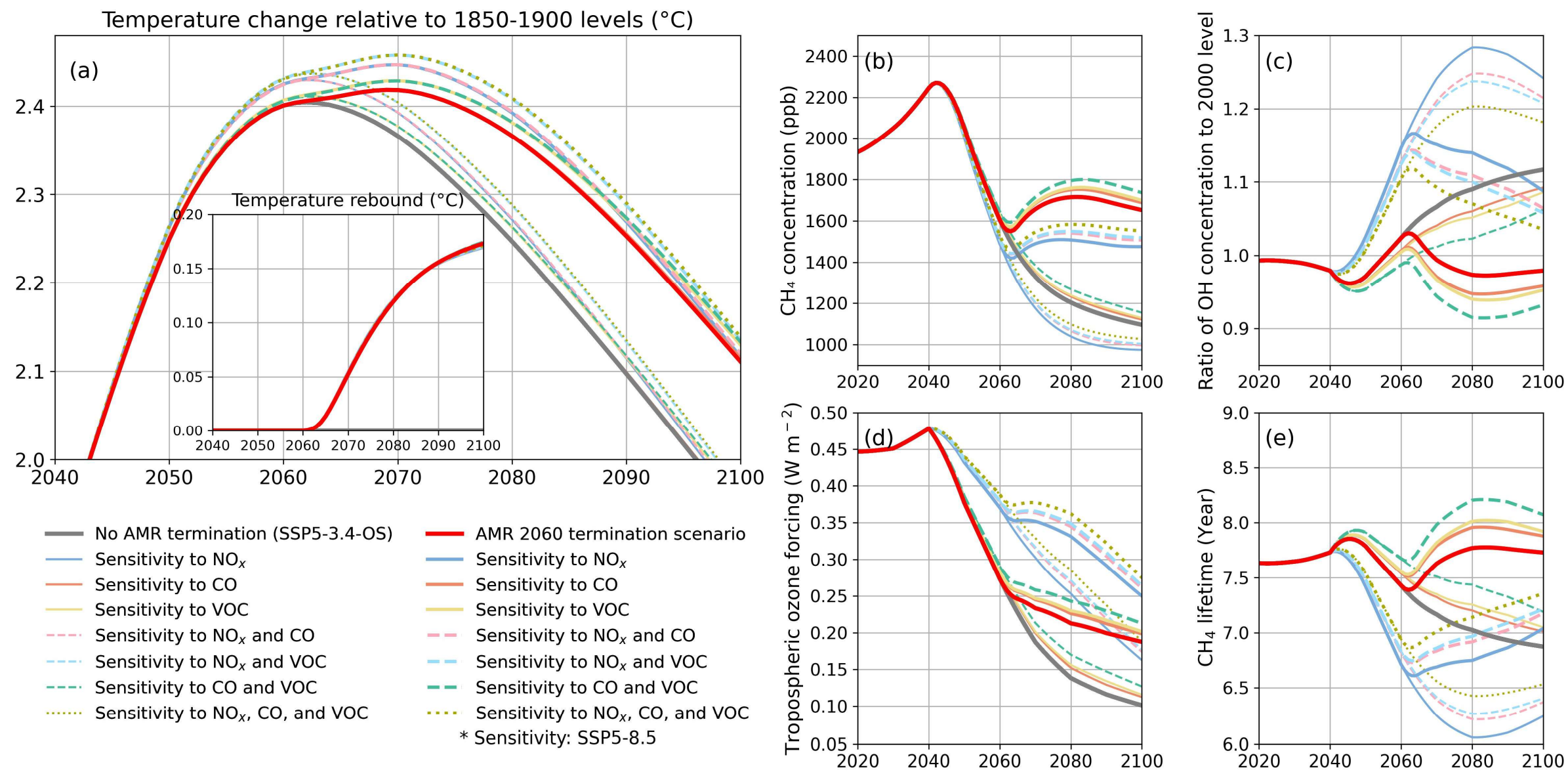

**Fig. S3 | Sensitivity of scenarios with and without AMR termination to background pollutant emissions at SSP5-8.5.** (a) Global-mean temperature change relative to 1850–1900 levels (°C) with inset showing temperature rebound (°C); (b) Atmospheric $CH_4$ concentration (ppb); (c) Ratio of OH concentration to 2000 levels; (d) Tropospheric ozone forcing (W/m$^2$); (e) Atmospheric $CH_4$ lifetime (yr). This figure complements Fig. S2 by extending the sensitivity analysis to SSP5-8.5. See the caption for Fig. S2.

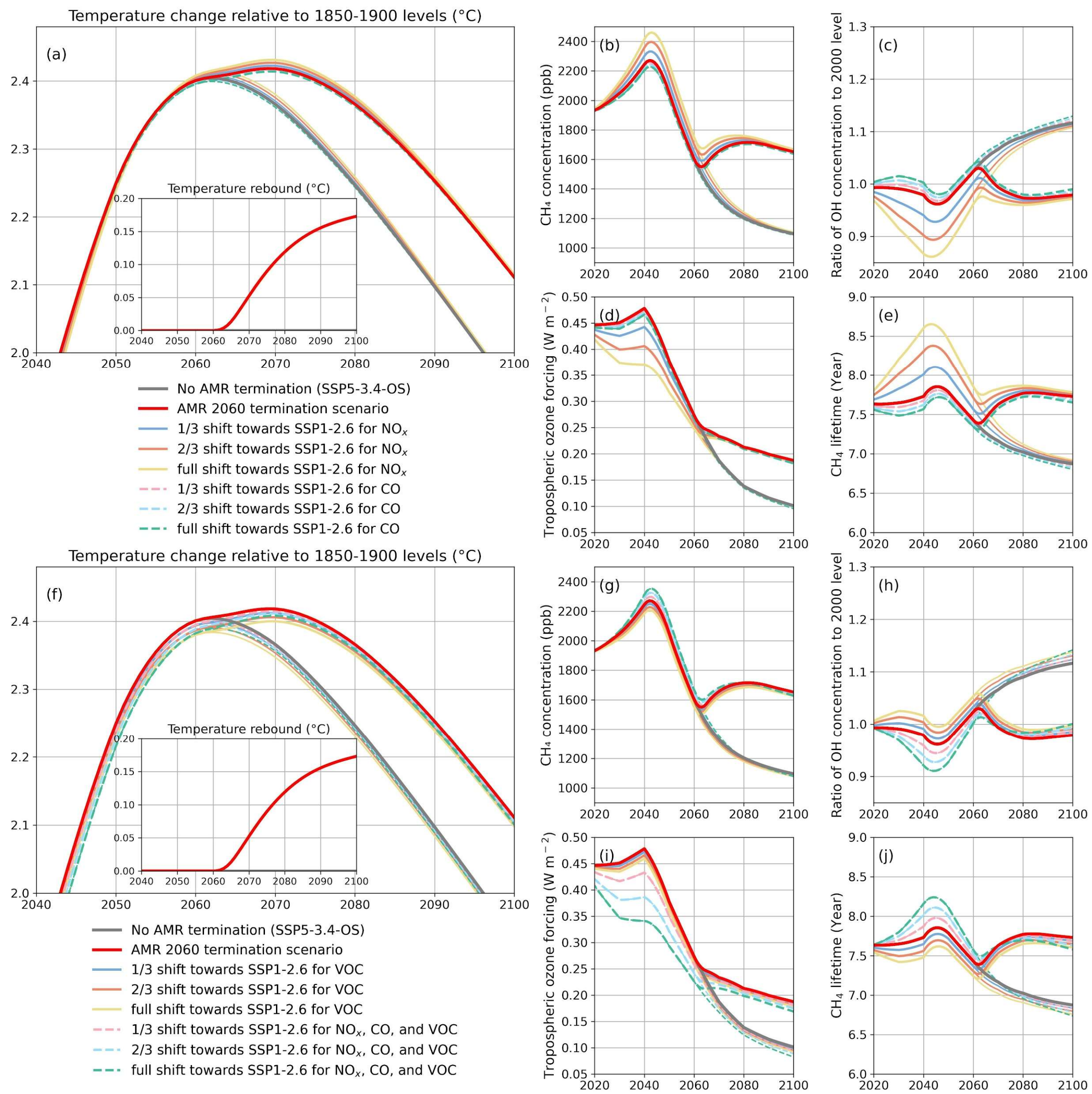

**Fig. S4 | Sensitivity of scenarios with and without AMR termination to intermediate background pollutant emissions.** (a) and (f) Global-mean temperature change relative to 1850–1900 levels (°C) with inset showing temperature rebound (°C); (b) and (g) Atmospheric $CH_4$ concentration (ppb); (c) and (h) Ratio of OH concentration to 2000 levels; (d) and (i) Tropospheric ozone forcing ($W/m^2$); (e) and (j) Atmospheric $CH_4$ lifetime (yr). This figure complements Fig. 2 and Fig. S2 by extending the sensitivity analysis to include interpolated scenarios between SSP5-3.4-OS and SSP1-2.6 (Fig. S1). It illustrates the sensitivity of the scenarios with and without AMR termination (red and gray lines, respectively) to variations in pollutant emissions ($NO_x$, CO, and VOC). While pollutant emissions follow SSP5-3.4-OS by default, the sensitivity cases assume the indicated pollutant shifts partially or fully toward SSP1-2.6.

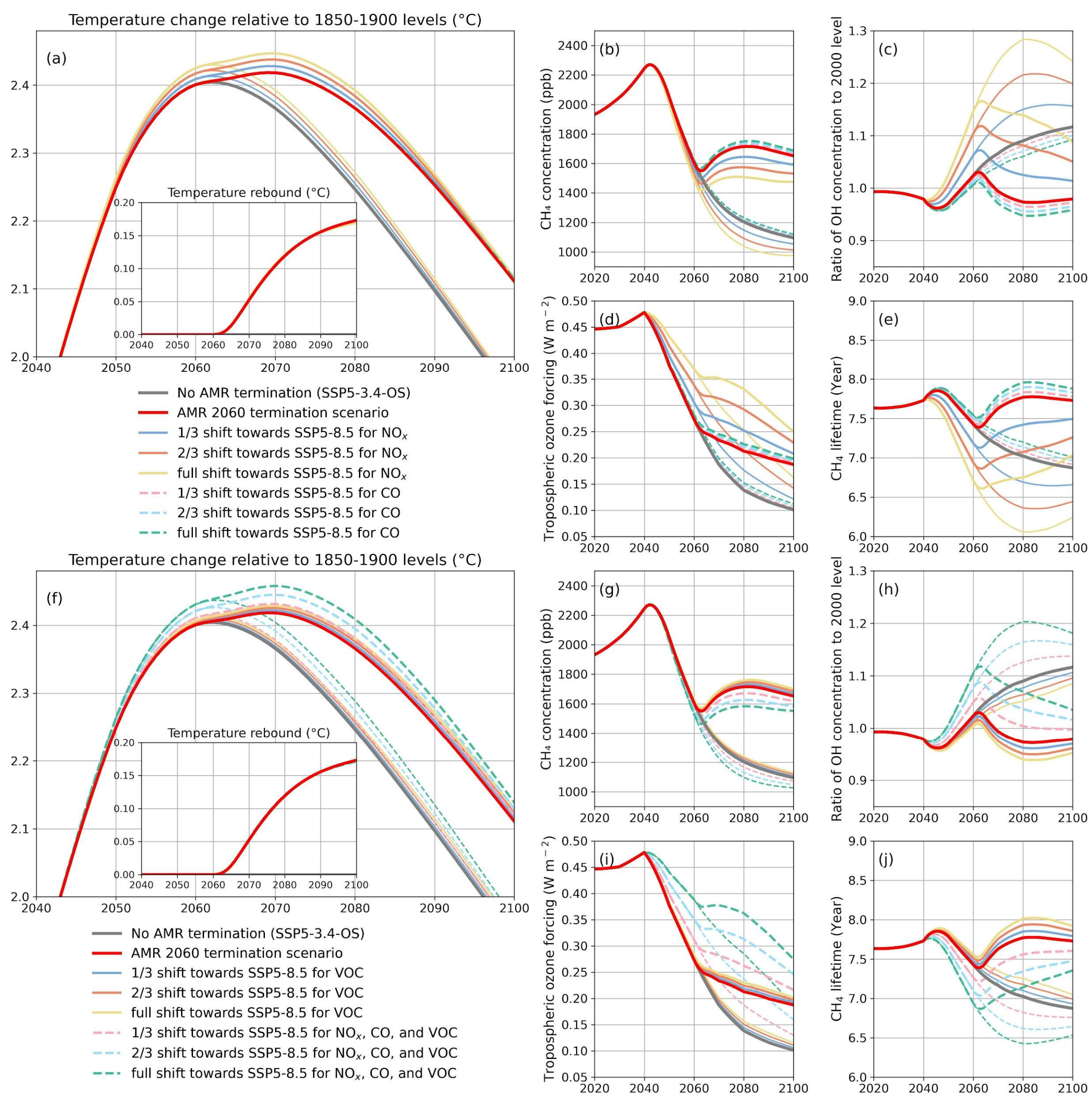

**Fig. S5 | Sensitivity of scenarios with and without AMR termination to intermediate background pollutant emissions.** (a) and (f) Global-mean temperature change relative to 1850–1900 levels (°C) with inset showing temperature rebound (°C); (b) and (g) Atmospheric $CH_4$ concentration (ppb); (c) and (h) Ratio of OH concentration to 2000 levels; (d) and (i) Tropospheric ozone forcing (W/m$^2$); (e) and (j) Atmospheric $CH_4$ lifetime (yr). This figure complements Fig. 2 and Fig. S3 by extending the sensitivity analysis to include interpolated scenarios between SSP5-3.4-OS and SSP5-8.5 (Fig. S1). See the caption of Fig. S4.

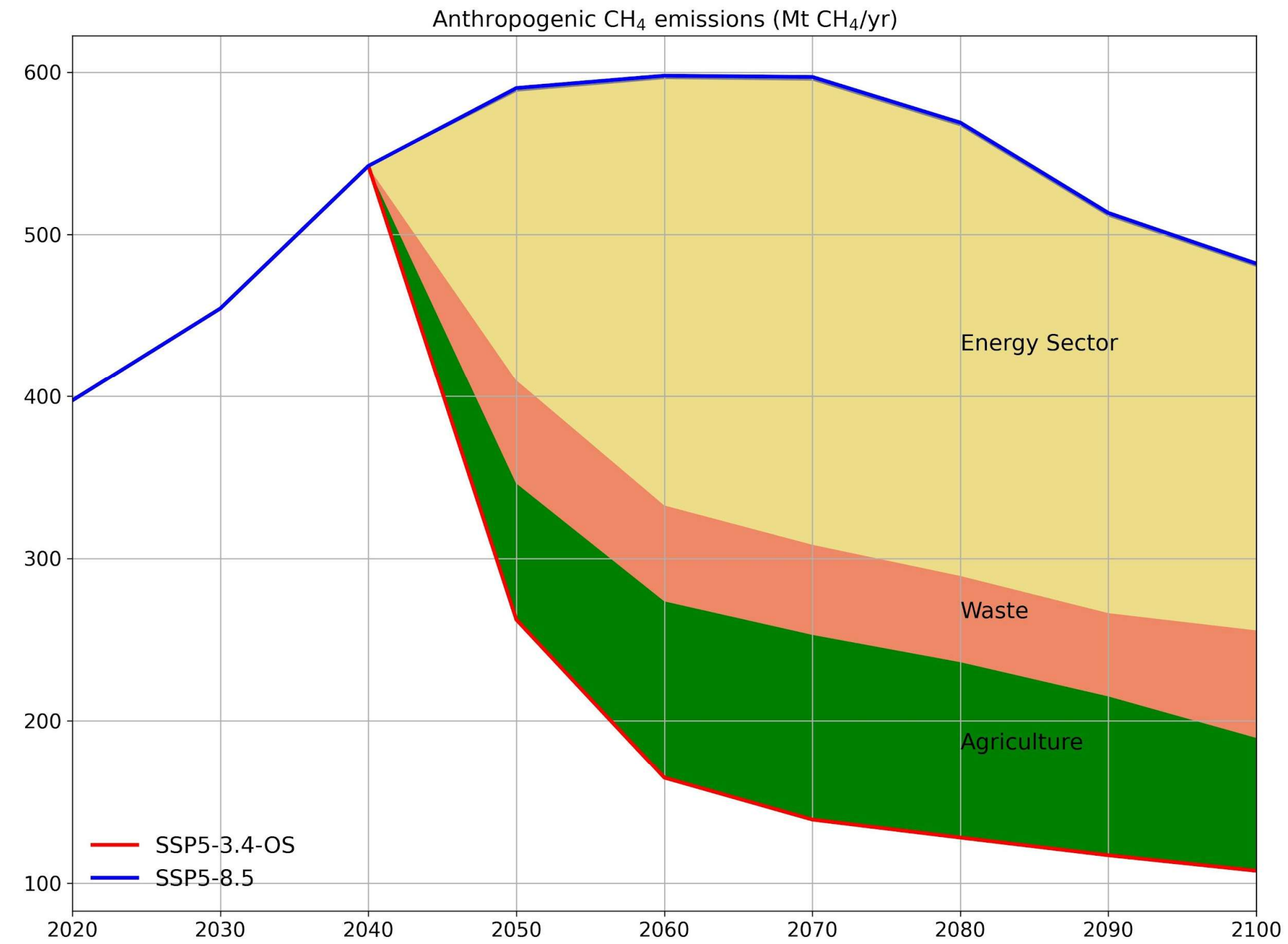

**Fig. S6 | Methane emissions from three major sectors (energy, agriculture, and waste) in SSP5-3.4-OS and SSP5-8.5.** The figure shows the sectoral emission differences between SSP5-3.4-OS and SSP5-8.5. For all other sectors, the difference between the two scenarios is negligible (< 1 Mt$CH_4$/yr). Data are derived from the IIASA SSP Database v2.0. In developing our climate intervention scenarios, we assumed that the emission reductions from the waste and agriculture sectors (red and green shaded areas, respectively) are achieved through AMR. In contrast, those from the energy sector (yellow shaded area) are attributed to conventional emission abatement measures rather than AMR.

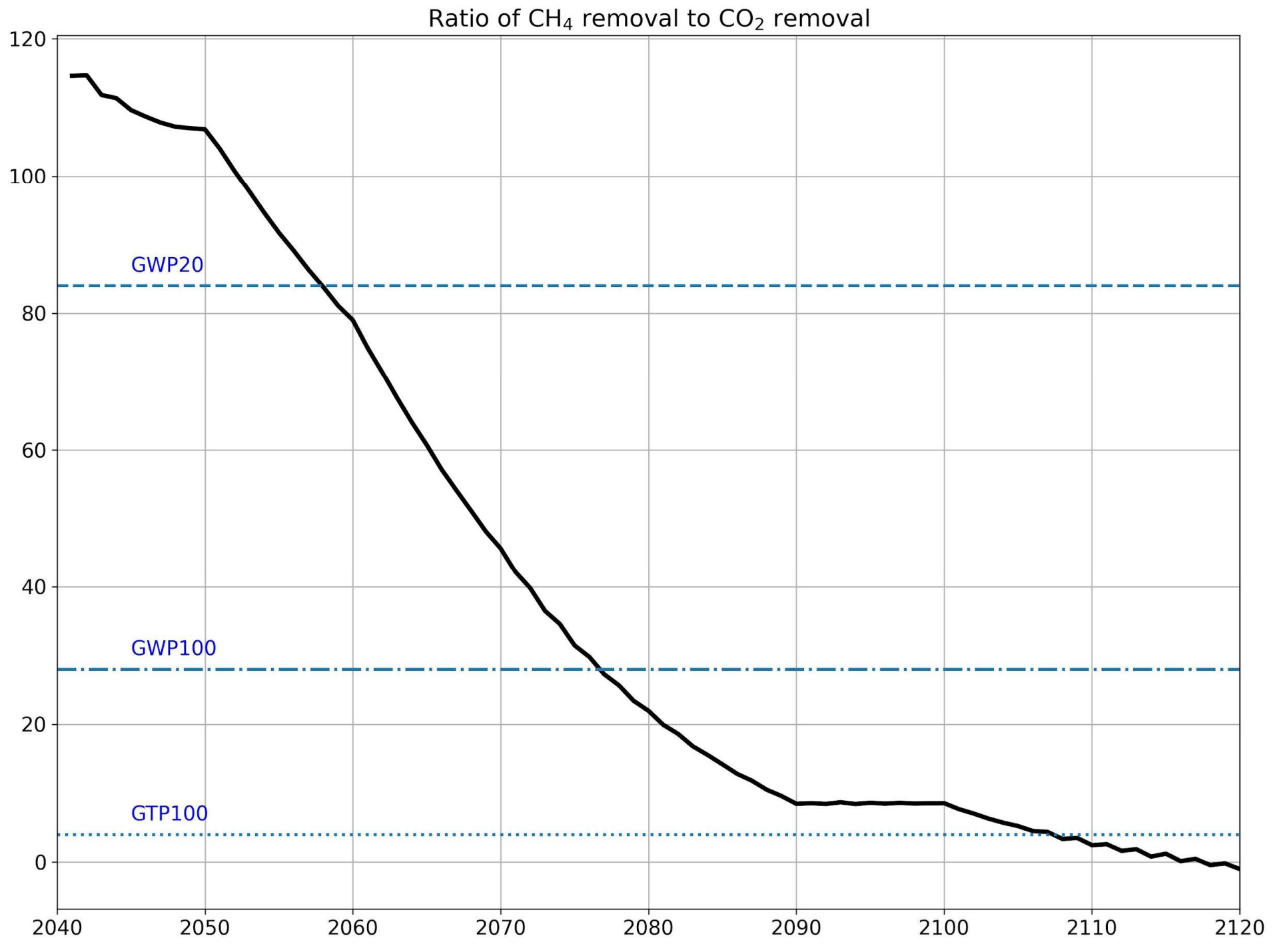

**Fig. S7 | Ratio of $CH_4$ removal to $CO_2$ removal based on the forcing equivalence method.** GWP20, GWP100, and GTP100 are shown for reference. See Online Methods for discussion.

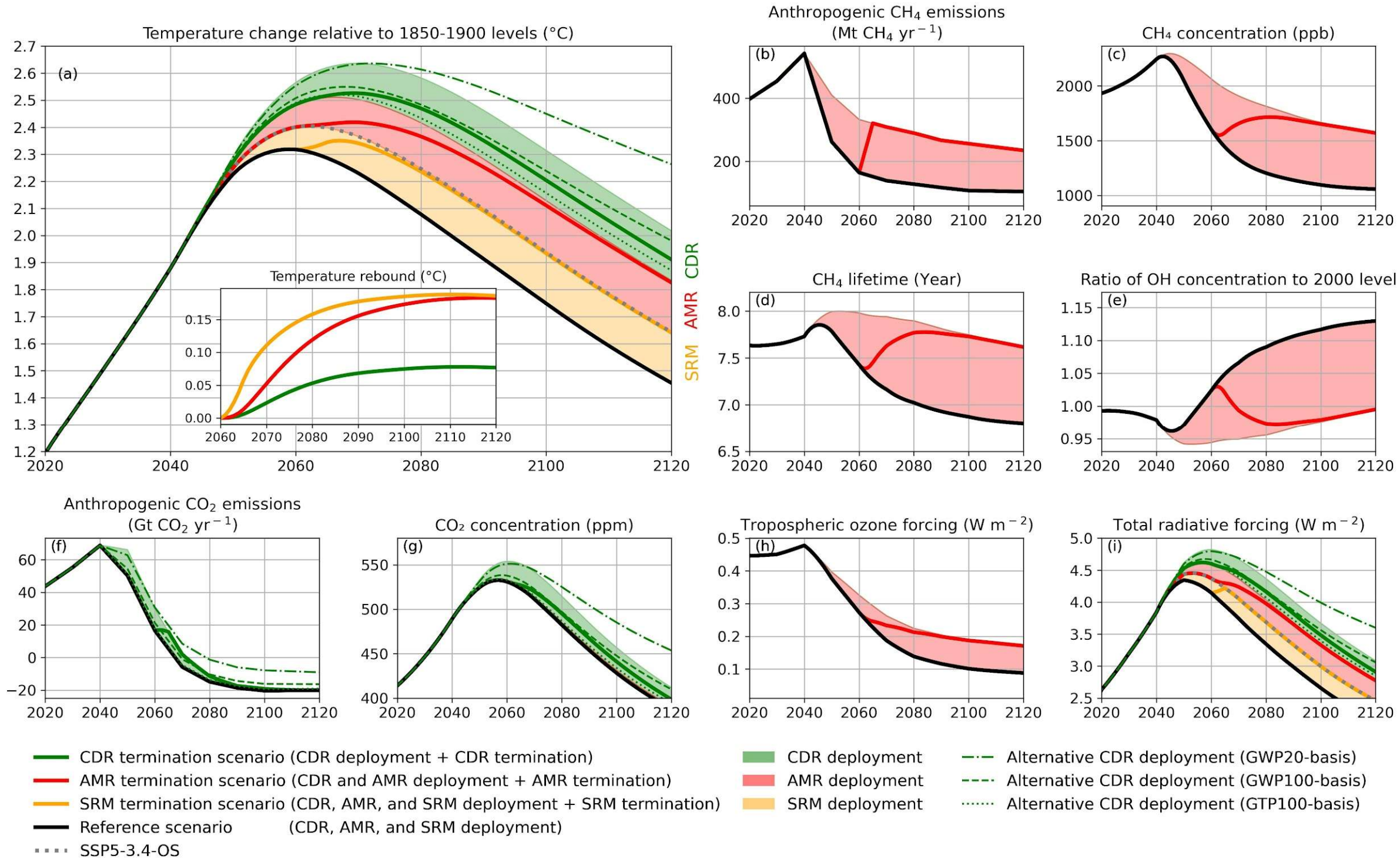

**Fig. S8 | Deployments and terminations of AMR, CDR, and SRM, including alternative CDR deployments.** (a) Global-mean temperature change relative to 1850–1900 levels (°C) with inset showing temperature rebound (°C); (b) Anthropogenic $CH_4$ emissions ($MtCH_4$/yr); (c) Atmospheric $CH_4$ concentration (ppb); (d) Atmospheric $CH_4$ lifetime (yr); (e) Ratio of OH concentration to 2000 levels; (f) Anthropogenic $CO_2$ emissions ($GtCO_2$/yr); (g) Atmospheric $CO_2$ concentration (ppm); (h) Tropospheric ozone forcing (W/m$^2$); (i) Total radiative forcing (W/m$^2$). This figure complements Fig. 1 by displaying projections for the alternative CDR deployments (Online Methods). Green dashed-dotted, dashed, and dotted lines represent such CDR deployments based on emission equivalence using GWP20, GWP100, and GTP100, respectively. Terminations are not simulated for alternative CDR deployments. For the default AMR, CDR, and SRM deployments, only the terminations beginning in 2060 are shown in thick red, green, and yellow lines, respectively.